\documentstyle[amsfonts,amssymb]{ioscrcb}

\newcommand{\Names}{{\cal N}}
\newcommand{\hd}{\mathsf{hd}}
\newcommand{\Conames}{\bar{{\cal N}}}
\newcommand{\barb}{\bar{(\ )}}
\newcommand{\Lab}{{\cal L}}
\newcommand{\Act}{{\mathsf Act}}
\newcommand{\Actne}{{\mathsf Act}_{+}}
\newcommand{\Actnest}{{\mathsf Act}_{+}^{\star}}
\newcommand{\pcomp}{\mid}
\newcommand{\rec}[2]{\mathsf{rec} \, #1 . \, #2} 
\newcommand{\labarrow}[1]{\stackrel{#1}{\longrightarrow}}
\newcommand{\olabarrow}[1]{\stackrel{#1}{\Longrightarrow}}
\newcommand{\nolabarrow}[1]{\stackrel{#1}{\not\!\Longrightarrow}}
\newcommand{\restrict}{{\setminus}}
\newcommand{\dunion}{\dot{\; \cup \;}}
\newcommand{\ddunion}{\dot{\cup}}
\newcommand{\conj}{\; \wedge \;}
\newcommand{\equivdef}{\; \equiv \;}
\newcommand{\eqdef}{\; \stackrel{\Delta}{=} \;}
\newcommand{\iffdef}{\;\; \stackrel{\Delta}{\Longleftrightarrow} \;\;}
\newcommand{\Failures}{{\mathcal F}}
\newcommand{\faileq}{\approx_{f}}
\newcommand{\seq}{\vdash}
\newcommand{\leftbank}{{\mathtt l}}
\newcommand{\rightbank}{{\mathtt r}}
\newcommand{\lapp}[2]{\langle #1 | #2}
\newcommand{\rapp}[2]{#1 | #2 \rangle}
\newcommand{\Implies}{\; \Longrightarrow \;}
\newcommand{\Tensor}{\otimes}

\newcount\PLv\newcount\PLw\newcount\PLx\newcount\PLy\newdimen\PLyy\newdimen\PLX
\newbox\PLdot \setbox\PLdot\hbox{\tiny.} \def\scl{.08} 
\def\PLot#1{\PLx`#1\advance\PLx-42\PLy\PLx\PLv\PLx\divide\PLy9\PLw\PLy\multiply
\PLw9\advance\PLx-\PLw\advance\PLx-4\PLy-\PLy\advance\PLy4\PLX=\the\PLx pt
\advance\PLyy\the\PLy pt\wd\PLdot=\scl\PLX\raise\scl\PLyy\copy\PLdot}
\def\draw#1{\ifx#1\end\let\next=\relax\else\PLot#1\let\next=\draw\fi\next}

\def\invamp{\hbox{\PLyy=70pt\draw :::;DMV_gqppyyyyyooooxxxnnwvlutkjaWNE=5-./9%
9:::CCCC:::99/..--544=EENWWaajjjkktttttttNNNVVVVVVVV\end \hskip8pt}}
\newbox\iabox\setbox\iabox\invamp 
\newcommand{\Par}{\invamp}
\newcommand{\With}{\&}
\newcommand{\Plus}{\oplus}
\newcommand{\Bang}{!}
\newcommand{\Whynot}{?}
\newcommand{\wbisim}{\faileq}
\newcommand{\varempty}{\varnothing}
\newcommand{\posforces}{\Vdash_{+}}
\newcommand{\negforces}{\Vdash_{-}}

\newcommand{\linimpl}{\multimap}

\newcommand{\isoarrow}{\stackrel{\cong}{\longrightarrow}}

\newcommand{\Cat}{\mathcal{C}}
\newcommand{\Ob}{\mathsf{Ob}}
\newcommand{\Val}{{\mathcal V}}

\newcommand{\Nat}{\omega}

\newcommand{\converges}{{\Downarrow}}
\newcommand{\diverges}{{\Uparrow}}

\newlength{\sqpreordheight}
\newlength{\sqpreorddepth}
\newcommand{\sqpreord}%
{\mathbin{\raisebox{-1.02ex}[\sqpreordheight][\sqpreorddepth]%
{$\stackrel{\textstyle \sqsubset}{\sim}$}}}
\newcommand{\sqgtpreord}%
{\mathbin{\raisebox{-1.02ex}[\sqpreordheight][\sqpreorddepth]%
{$\stackrel{\textstyle \sqsupset}{\sim}$}}}

\newcommand{\lsem}{[\! [}

\newcommand{\rsem}{]\! ]}

\newtheorem{proposition}{Proposition}[section]

\title{Process Realizability}
\author{Samson Abramsky}
\date{}
\begin{document}

\bibliographystyle{alpha}
\maketitle
\section{Introduction}
Realizability has proved to be a fruitful approach to the semantics of 
computation, see e.g. \cite{AL,Cro,Lon,AC}.
The \emph{scope} of realizability methods has been limited to
\emph{Intuitionistic Logic} (with some extensions to Classical Logic),
on the logical side, and to \emph{functional computation} on the computational
side. Our aim in the present work is to explore the possibilities  for
broadening the scope of realizability:
\begin{itemize}
\item beyond Intuitionistic Logic, to Classical Linear Logic, and more;
\item beyond functional computation, to encompass concurrent and 
non-deterministic computation.
\end{itemize}
Why do this? We shall mention just one, fairly concrete motivation.
Consider the well-established paradigm of extracting \emph{functional programs}
from (Intuitionistic or Classical) \emph{proofs}, using the Curry-Howard
isomorphism or realizability \cite{GLT,Sch}. Can we analogously find a suitable
combination of a logic and a realizability universe such that we can extract
interesting \emph{concurrent programs}---communication protocols,
distributed algorithms, security protocols---from proofs
of their specifications?

Two important caveats should be registered here. The first is that we don't 
envisage the extraction of programs from proofs as a practical programming 
methodology. However, in the case of functional computation, the 
well-understood paradigm of program extraction from proofs
is a key component of our
foundational understanding of functional programming; the objective here
is to attain a comparably well-founded paradigm for concurrent programming.
The second caveat is that we don't---as yet---claim to be able to extract
\emph{interesting} concurrent programs, in the above sense, from proofs.
However, we \emph{do} see the ideas which we shall now put forward as a step
in this direction. 
\paragraph{Note to the reader}
This paper aims to give a readable and reasonably accessible account
of some ideas linking the currently still largely separate worlds
of concurrency theory and process algebra, on the one hand, and
type theory, categorical models and realizability on the other.
Background in process algebra may be found in standard texts such as
\cite{Hoa85,Hen,Mil89,Ros}; while background in realizability, 
categorical models etc.
is provided by texts such as \cite{GLT,AL,Cro,AC,BW}.
A modest background in either or both of these fields should suffice
to understand the main ideas.
Most of the detailed verification of properties of the  formal definitions
we will present is left as a series of exercises. The diligent
reader who attempts a number of these should get some feeling for the
interplay between concrete process-theoretic notions, and more abstract
logical and categorical ideas, which is characteristic of this topic.
It is this interplay which makes the topic a fascinating one for the
author; I hope this brief introduction, to a field which is
still wide open for further development,  succeeds in conveying something
of this fascination to the reader.

\section{CCS with simultaneous actions}
Our universe of realizers will be a minor extension of one of the most
standard and widely-used process calculi, namely Milner's CCS \cite{Mil89}.
The extension is to allow ``compound'' actions, consisting of the simultaneous
performance of several ``atomic'' actions. This idea of compound actions
is present in the synchronous process calculus SCCS \cite{Mil89}; the point
here is to introduce this as an extension of the asynchronous calculus
CCS. Our reason for using this extension is that it will allow us to
realize identities and other canonical isomorphisms as ``wires'',
with typical behaviour
\[ \alpha \; \underline{\ \ \ \ \ \ \ \ }  \; \beta \]
in which two signals are propagated simultaneously, at the two ``ends of
the wire'', so to speak. This extension is not new; it was introduced in the 
present author's work on asynchronous interaction categories 
\cite{Abr,AGN96},  for
similar reasons. Much of what we will do here can be seen as a recasting
of the work on interaction categories into a realizability framework.
Indeed, the essential ideas on the process interpretation of proofs
go back to a 1991 lecture on ``Proofs as Processes'' (see \cite{Abr94}).

\subsection{Names, co-names and actions}
As usual with CCS, we introduce two disjoint, countable sets $\Names$ of
\emph{names}, and $\Conames$ of \emph{co-names}, with a bijection
$\barb : \Names \isoarrow \Conames$, which we extend to an involution
\[ \barb : \Names + \Conames \isoarrow \Conames + \Names . \]
We use $\alpha$, $\beta$, $\gamma$ to range over names, and write 
$\bar{\alpha}$, $\bar{\beta}$, $\bar{\gamma}$ for the corresponding co-names,
with $\bar{\bar{\alpha}} = \alpha$ etc.
We write $\Lab = \Names \cup \Conames$ for the set of \emph{labels},
ranged over by $\lambda$, $\mu$, $\nu$.
An \emph{action} will be a finite set of labels. We write $\Act$ for the
set of actions, ranged over by $a$, $b$, $c$. In particular, we write
$\tau$ for the empty set $\varempty$. The interpretation of an action
$a = \{ \lambda_1 , \ldots , \lambda_n \}$ is that the simultaneous
performance of the actions in $a$ is observed; thus $\tau$ can be viewed
as a ``silent'' or ``unobservable'' or ``internal'' action.
The involution $\barb$ is extended pointwise to actions:
\[ \bar{a} = \{ \bar{\lambda} \mid \lambda \in a \} . \]
Note that $\bar{\bar{a}} = a$, and that $\bar{\tau} = \tau$.

\subsection{Guarded Terms}
We now introduce a class of \emph{guarded} process terms, with the following 
syntax.

\[ 
P \;\; ::= \;\;  a.P \, \mid \, \Sigma_{i \in I} a_{i}.P_i 
\; (\forall i \in I. \, a_i \neq \tau ) 
\, \mid \, P \pcomp Q \, \mid \, G \restrict L  \, \mid \, G[f] 
\, \mid \, X \, \mid \, \rec{X}{P} .
\]
Here $I$ ranges over countable index sets,
$L$ ranges over subsets of $\Lab$ (\textit{i.e.} \emph{sorts}),
$X$ ranges over a set of process variables, and $f$ over \emph{renamings},
i.e. $\barb$-preserving injective functions on $\Lab$. (In fact, we shall
allow \emph{partial} injective functions as renamings, with the proviso
that the \emph{sort} of the process to which the renaming is applied
is contained in the domain of the function. For details---which are easy and
standard---see \cite{Mil89}). Renamings are extended pointwise to actions.
As usual, the empty sum is written as $0$, and the binary case as
$a.P + b.Q$.

\subsection{Transitions}
We define the labelled transitions $\labarrow{a}$, ($a \in \Act$) by the
following inductive definition.

\[ \frac{}{a.P \labarrow{a} P} \qquad \qquad
\frac{}{\Sigma_{i \in I} a_{i}.P_i \labarrow{a_j} P_j \;\; (j \in I) }\]

\[ \frac{P \labarrow{a} Q}{P \restrict L \labarrow{a} Q \restrict L} \;\;
(a \cap (L \cup \bar{L}) = \varempty ) \qquad \qquad
\frac{P \labarrow{a} Q}{P[f] \labarrow{f(a)} Q[f]} \]

\[ \frac{P[\rec{X}{P}/X] \labarrow{a} Q}{\rec{X}{P} \labarrow{a} Q} \]

\[ \frac{P \labarrow{a} P'}{P \pcomp Q \labarrow{a} P' \pcomp Q} \qquad \qquad
\frac{Q \labarrow{a} Q'}{P \pcomp Q \labarrow{a} P \pcomp Q'} \]

\[ \frac{P \labarrow{a \ddunion b} P' \;\; Q \labarrow{\bar{b} \ddunion c} 
Q'}{P
\pcomp Q \labarrow{a \ddunion c} P' \pcomp Q'} \]
Here $a \dunion b$ means that $a$ and $b$ are disjoint; this, together with 
$b \dunion c$ and $a \dunion c$ should be viewed
as \emph{constraints} on the applicability of the rule.
These rules are completely standard, except for the last, which
generalizes the usual rule
\[ \frac{P \labarrow{\lambda} P' \;\; Q \labarrow{\bar{\lambda}} Q'}{P \pcomp Q
\labarrow{\tau} P' \pcomp Q'} \]
(Take $a = c = \varempty$, $b = \{\lambda \}$).

Note that we can define the wire $W_{\lambda , \mu}$ which repeatedly
performs the compound action
\[ \lambda \; \underline{\ \ \ \ \ \ \ \ } \; \beta \]
as the term
\[ W_{\lambda , \mu} \eqdef \rec{X}{\{\lambda , \mu \} . X} . \]

\subsection{Failures Equivalence}
Let $\Actne$ be the set of non-empty actions.
We define the observable transition relations $\olabarrow{a}$, for
$a \in \Actne$, as $\labarrow{\tau}^{\star} \labarrow{a} 
\labarrow{\tau}^{\star}$, and extend this to $\olabarrow{s}$ for strings
$s \in \Actnest$ in the usual fashion.
We define the set of 
\emph{failures} of a process $P$ by
\[ \Failures (P) = \{ (s, X) \mid \ s \in \Actnest \conj X \subseteq \Actne
\conj \exists Q. \, (P \olabarrow{s} Q \conj
\forall a \in X. \, Q \nolabarrow{a} ) \} . \]
We define failures equivalence by
\[ P \faileq Q \iffdef \Failures (P) =
\Failures (Q) . \]
\begin{proposition}
Failures equivalence is a congruence on guarded terms.
\end{proposition}
\paragraph{Discussion}
Our reason for working with failures equivalence (\cite{BHR,BR83}) is that it, 
or
one of its variants such as the testing equivalences of Hennessy and De Nicola
\cite{DNH}, or the Failures-Divergences model of Brookes and Roscoe \cite{BR84},
seem to be the finest equivalences which will suffice for our purposes.
In particular, the realizability for the additive connectives of Linear
Logic will not work in a fully satisfactory way if we use a finer
equivalence such as weak congruence or weak bisimulation.
It is worth noting that failures equivalence (or more accurately,
the refined Failures-Divergences equivalence) is the standard equivalence
for CSP, the other widely used process calculus.

Our reason for using guarded sums is to give a slightly simplified treatment
of non-determinism. We could equally well have introduced the usual
external and internal choice constructs as in \cite{Hoa85,Hen,Ros}.

The representation of processes in terms of their failures
gives rise to a fully abstract model for failures equivalence, on which
the process operations can be defined in a denotational, compositional
fashion \cite{BHR,Ros}. We have presented the semantics in an operational style
to be concrete and simple, but in many ways the denotational presentation
is more elegant and illuminating. 
It is also worth noting that we could equally well work with
CSP as our process calculus rather than CCS, using the same underlying
denotational model.
\subsection{Some basic combinators}
Since processes are untyped, we will build a type-free universe of
realizers. As usual, this will require a little coding (\textit{cf.} the
Kleene algebra $K_0$ \cite{Kle}, graph models \cite{Sco76} etc.), but in our
setting this will take a very simple form. We simply split the set of
names into two infinite disjoint sets
\[ \Names = \Names_l \dunion \Names_r \]
and fix bijections
\[ \leftbank : \Names \isoarrow \Names_l \;\;\;\;\quad 
\rightbank : \Names \isoarrow 
\Names_r \]
and extend  these in a $\barb$-respecting way to the set of labels:
\[ \Lab = \Lab_l \dunion \Lab_r \]
\[ \leftbank : \Lab \isoarrow \Lab_l \quad \;\;\;\; \rightbank : \Lab \isoarrow \Lab_r . \]
This means that we can view an arbitrary process $P$ as having its interface
to its environment split into two disjoint parts:
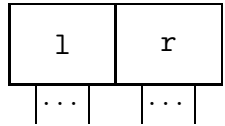
\begin{center}
\begin{picture}(100,50)(0,-15)
\put(8,0){\framebox(40,30){$\leftbank$}}
\put(48,0){\framebox(40,30){$\rightbank$}}
\put(18,0){\line(0,-1){15}}
\put(28,-7.5){\makebox(0,0){\ldots}}
\put(38,0){\line(0,-1){15}}
\put(58,0){\line(0,-1){15}}
\put(68,-7.5){\makebox(0,0){\ldots}}
\put(78,0){\line(0,-1){15}}
\end{picture}
\end{center}
We can use this splitting of the name space to define some combinators
which will play a fundamental role in our notion of process realizability.

Firstly, we have a tensor product which will express \emph{disjoint
(non-communicating) parallel composition}:
\[ P \otimes Q \eqdef P[\leftbank ] \pcomp Q[\rightbank ] . \]
The use of the relabelling functions \emph{forces}  the two processes to be 
disjoint:
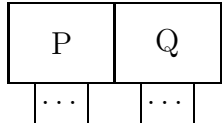
\begin{center}
\begin{picture}(100,50)(0,-15)
\put(8,0){\framebox(40,30){P}}
\put(48,0){\framebox(40,30){Q}}
\put(18,0){\line(0,-1){15}}
\put(28,-7.5){\makebox(0,0){\ldots}}
\put(38,0){\line(0,-1){15}}
\put(58,0){\line(0,-1){15}}
\put(68,-7.5){\makebox(0,0){\ldots}}
\put(78,0){\line(0,-1){15}}
\end{picture}
\end{center}
Secondly, we can define a notion of \emph{application}. As usual in a type-free
setting, we have to be able to see arbitrary elements of our universe
either as ``functions'' or as ``arguments'', as required. The basic
splitting of our name space
allows us to see $P$ as a ``function'' in \emph{two, entirely symmetrical
ways}. We can see the left part of the name space as the ``attachment point''
for an argument, with the right part left free to communicate the
``result''; or we can attach on the right and transmit the result through
the left.

The first view leads to a \emph{left} (or \emph{forwards}) application:
\[ \lapp{Q}{P} \eqdef ((Q[\leftbank ] \pcomp P) \restrict \Lab_l )[\rightbank^{-1}] \]
which we can visualize as follows:
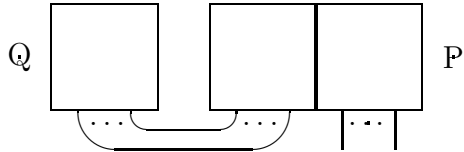
\begin{center}
\begin{picture}(100,60)(30,-15)
\put(-4,20){\framebox(0,0){Q}}
\put(8,0){\framebox(40,40){}}
\put(68,0){\framebox(40,40){}}
\put(108,0){\framebox(40,40){}}
\put(160,20){\framebox(0,0){P}}
\put(58,0){\oval(80,30)[b]}
\put(58,0){\oval(40,15)[b]}
\put(30,-5){\makebox(0,0){\ldots}}
\put(88,-5){\makebox(0,0){\ldots}}
\put(118,0){\line(0,-1){15}}
\put(128,-5){\framebox(0,0){\ldots}}
\put(138,0){\line(0,-1){15}}
\end{picture}
\end{center}
We use a Dirac-style bra-ket notation (\textit{cf.} \cite{Dir}) to denote the
left application of $P$ to $Q$. The idea is that $Q$ is relabelled
into the left part of $P$'s name space, and we then restrict on the left
name space  so that $P$ and $Q$ are forced to interact there; the
resulting observable behaviour is purely that produced by $P$ in the right
part of its name space. Finally, we ``normalize'' by relabelling 
back into the global name space, using the inverse of the bijection 
$\rightbank$.

Symmetrically, we can define a \emph{right} (or \emph{reverse}) application:
\[ \rapp{P}{R} \eqdef ((P \pcomp R[\rightbank ]) \restrict \Lab_r )[\leftbank^{-1}] \]

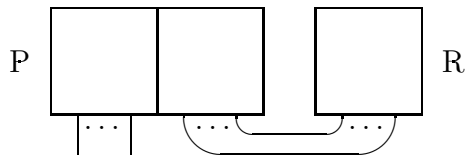
\begin{center}
\begin{picture}(100,60)(-10,-15)
\put(-44,20){\framebox(0,0){P}}
\put(-32,0){\framebox(40,40){}}
\put(8,0){\framebox(40,40){}}
\put(68,0){\framebox(40,40){}}
\put(120,20){\framebox(0,0){R}}
\put(-22,0){\line(0,-1){15}}
\put(-12,-5){\makebox(0,0){\ldots}}
\put(-2,0){\line(0,-1){15}}
\put(58,0){\oval(80,30)[b]}
\put(58,0){\oval(40,15)[b]}
\put(30,-5){\makebox(0,0){\ldots}}
\put(88,-5){\makebox(0,0){\ldots}}
\end{picture}
\end{center}
\section{Realizability}
We will now define a notion of realizability for formulas built from the
propositional connectives of (Classical) Linear Logic: the multiplicatives
$\Tensor$ and $\Par$, the additives $\With$ and $\Plus$, and the
exponentials $\Bang$ and $\Whynot$. More precisely, we shall define
two relations
\[ P \posforces A \;\;\; \mbox{and} \;\;\; P \negforces A \]
between process terms $P$ and formulas $A$, by induction on the construction
of $A$. We shall read $P \posforces A$ as: ``$P$ is a realizer/strategy/proof
of $A$'', or ``$P$ is a value of type $A$'';   and $P \negforces A$ as:
``$P$ is a counter-realizer/counter-strategy/refutation of $A$'', or
``$P$ is an $A$-consuming context''. This builds a classical (involutive)
duality, and specifically an interpretation of the Linear negation
$(\_)^{\bot}$, into our realizability interpretation. In particular, we will
have:
\[ P \posforces A^{\bot} \equivdef P \negforces A \]
\[ P \negforces A^{\bot} \equivdef P \posforces A . \]

\subsection{Multiplicatives}
For each connective, we must define both the positive and negative
notions of realizability. However, once this is done for one connective,
the notions for the de Morgan duals are also determined.

\noindent The definitions for the tensor product are as follows:
\[ P \posforces A \otimes B \equivdef P \wbisim P_1 \otimes P_2 \conj 
P_1 \posforces A \conj P_2 \posforces B . \]
Note that, because of the disjoint relabelling in the definition of
the tensor combinator on processes:
\[ P_1 \otimes P_2 \wbisim Q_1 \otimes Q_2 \Implies P_1 \wbisim Q_1
\conj P_2 \wbisim Q_2 \]
so the decomposition in the above clause is unique up to failures equivalence
(in fact, up to weak bisimulation).

\[ P \negforces A \otimes B \equivdef \forall Q. \, (Q \posforces A 
\, \Rightarrow \, \lapp{Q}{P} \negforces B)
\conj
\forall R. \, (R \posforces B \Rightarrow \rapp{P}{R} \negforces A) . \]
This is a symmetrized version of the familiar ``logical relations''
or realizability condition. $P$ counter-realizes the multiplicative
conjunction $A \otimes B$ if it carries every realizer of $A$, under
forwards application, to
a counter-realizer for $B$, and every realizer of $B$, under reverse 
application, to a counter-realizer for $A$.

\noindent Applying de Morgan duality, this yields the more familiar-looking definition
for linear implication:
\[ A \linimpl B \eqdef A^{\bot} \Par B \eqdef (A \otimes B^{\bot})^{\bot} . \]
\[ P \posforces A \linimpl B \equivdef 
\forall Q. \, (Q \posforces A \, \Rightarrow \, \lapp{Q}{P} \posforces B)
\conj
\forall R. \, (R \negforces B \, \Rightarrow \, \rapp{P}{R} \negforces A) . \]
$P$ realizes the linear implication $A \linimpl B$ if it carries
realizers of $A$ to realizers of $B$, and counter-realizers of $B$
to counter-realizers of $A$.

The reading of the clause for negative realizability for the linear 
implication is also interesting:
\[ P \negforces A \linimpl B \equivdef P \wbisim Q \otimes R 
\conj Q \posforces A \conj R \negforces B . \]
This can be read as saying that $P$ realizes a context for consuming a
``linear function'' $f$ of type $A \linimpl B$ if it decomposes as an
input of type $A$ to be plugged into $f$, and a context of type $B$
for consuming the corresponding output.

It is interesting to note that the combinators $\otimes$, $\lapp{\cdot}{\cdot}$
and $\rapp{\cdot}{\cdot}$ which we introduced in order to realize the
multiplicative connectives are defined purely in terms of the
\emph{static} operators of CCS in Milner's classification \cite{Mil89},
namely parallel composition, restriction and relabelling. As we shall
now see, the additive connectives will be realized using only the
\emph{dynamic} operators of CCS (prefixing and summation), while the
exponentials will require a combination of the two, together with recursion.
\subsection{Additives}
Firstly, we fix once and for all two distinct names, say $\alpha$ and
$\beta$. These will be used to distinguish the left and right cases in the 
additive choice constructs $A \With B$ and $A \oplus B$. 
For the additive product or conjunction $A \With B$, the Opponent or
Environment will make the choice, and the realizers for $A \With B$
will have the form $\alpha . P + \beta . Q$, where $P$ is a realizer for
$A$ and $Q$ is a realizer for $B$. For the additive sum or disjunction
$A \oplus B$, Player or System makes the choice, and realizers for
$A \oplus B$ either have the form $\bar{\alpha}.R$, where $R$ is a realizer
for $A$, or $\bar{\beta}.S$, where $S$ is a realizer for $B$.

This leads to the following formal definitions:
\[ \begin{array}{lcl}
P \posforces A \With B & \equiv & P \faileq \alpha . Q + \beta . R 
\;  \conj \; Q \posforces A \; \conj \; R \posforces B  \\
P \negforces A\With B & \equiv & (P \faileq \bar{\alpha}.Q  \conj 
Q \negforces A) \;
\vee \;
(P \faileq \bar{\beta}.R  \conj  R \negforces B ) \\
P \posforces A \oplus B & \equiv & (P \faileq \bar{\alpha}.Q  \conj 
Q \posforces A) 
\; \vee \;
(P \faileq \bar{\beta}.R  \conj  R \posforces B) \\
P \negforces A \oplus B & \equiv  & P \faileq \alpha . Q + \beta . R 
\; \conj \; Q \negforces A \; \conj \; R \negforces B . 
\end{array} \]

\subsection{Exponentials}
As already indicated, to interpret the exponentials we combine additive and
multiplicative features with recursion on processes, and also induction to
define the realizability relation. This use of recursion and induction will
be set in a more general context in our later discussion of inductive
and coinductive types.

\noindent As for the additives, we fix some global names: $\omega$ 
(for weakening),
$\delta$ (dereliction) and $\gamma$ (contraction).
We define a process combinator
\[ !P \; \eqdef \; \rec{X}{\omega . 0 + \delta . P + \gamma . (X[\leftbank ] \pcomp
X[\rightbank ])} . \]
We can then define positive realizability for $!$:
\[ P \posforces \, !A  \equivdef  
P \faileq \, !Q \conj Q \posforces A . \]
We can think of $!P$ as a process which the environment can request to:
\begin{itemize}
\item deliver one copy of $P$ (dereliction)
\item be discarded (weakening)
\item make two copies of itself (contraction).
\end{itemize}
The negative realizability for $!A$ can be defined as follows.
\[ \begin{array}{ll}
P \negforces \, !A  \equivdef & (P \faileq \omega . 0) 
\; \vee \;
(P \faileq \bar{\delta}.Q \conj Q \negforces A ) \\
& \vee \;\;
(P \faileq \bar{\gamma}.Q \conj 
(\forall R \, \posforces A. \, \lapp{\, !R \,}{\, Q} \negforces \, !A 
\conj \rapp{Q \,}{\, !R} \, \negforces \, !A )).
\end{array} \]
Note that for the first time, the realizability relation is not being defined
purely by structural induction on the formula. Rather,
$P \negforces \, !A$ is being defined inductively, 
as the least fixed point
of the evident monotone operator on sets of processes which can be extracted
from the above definition, keeping $!A$ fixed. Note, however, that the
universal quantifier $\forall R \posforces A$ is ranging over realizers
for $A$, which we \emph{can} take to be already defined by structural
induction on $!A$. This is crucial for monotonicity.

The realizability relation for the dual connective $?$ is defined by De Morgan
duality from that for $!$, since $?A = (! A^{\bot})^{\bot}$. To understand
the inductive definition, think of it as defining the set of all $!A$-consuming
contexts, which request a number of copies of a realizer for $A$, and
then consume these copies. Recall that $?A$ is  the ``$\Par$-monoid
generated by $A$'', just as $!A$ is the ``$\otimes$-comonoid cogenerated
by $A$''. Thus the inductive clause for contraction parallels that for Par,
just as the case for contraction in the recursive definition of $!P$
parallels that for Tensor. This also says that there is no communication 
between the copies of $P$ in $!P$, while the counter-realizers  for $!A$ can 
use the information obtained from each copy in interacting with the others, as
for Par.
\subsection{Interpretation of Proofs}
We now show how to assign a realizer for $A$ to each proof of $A$
in Linear Logic. We will then be able to extract concurrent processes
as realizers from proofs.

We extend realizability to sequents $\Gamma = A_1 , \ldots , A_k$, treating
$\Gamma$ as $\Par \Gamma$ in the obvious fashion. We now indicate how
to assign realizers to sequent proofs in Linear Logic.

\subsubsection{Identity Axioms}
\[ \frac{}{\seq A^{\bot}, A} \]
All instances of the Identity Axioms are realized by the process
\[ I \eqdef \rec{X}{\sum_{a \in \Actne} (\leftbank (a) \cup 
\overline{\rightbank 
(a)}). X} \]
This behaves like a wire for any choice of actions
\[ a \; \underline{\ \ \ \ \ \ \ \ } \; \bar{a} \]
where the action at the left hand end of the wire corresponds to 
$\leftbank (a)$,
and that on the right to $\overline{\rightbank (a)} = \rightbank (\bar{a})$. To check that
$I$ is indeed a realizer for $A^{\bot} \Par A = A \linimpl A$ amounts
to verifying the process-algebraic fact that
\[ \forall P. \, \lapp{P}{I} \faileq P \faileq \rapp{I}{P} . \]
\paragraph{Exercise} Verify!

\subsubsection{Cut Rule}
\[ \frac{P \posforces \Gamma , A \quad \quad Q \posforces A^{\bot}, \Delta}{P ; Q
\posforces \Gamma , \Delta} \]
Here we are inductively assuming that we have already assigned a (positive)
realizer $P$ to the proof of $\Gamma, A$, and a realizer $Q$ to the
proof of $A^{\bot}, \Delta$. We must construct a realizer $P ; Q$ for
$\Gamma , \Delta$. This composition combinator will simultaneously generalize
the left and right application combinators we have previously introduced.

\noindent We fix a three-fold decomposition of the name space:
\[ \Names = \Names_1 \dunion \Names_2 \dunion \Names_3 \]
with bijections $\phi_{i,j} : \Names \isoarrow \Names_i \dunion \Names_j$
such that $\phi_{i,j}(\Names_{l}) = \Names_i$, $\phi_{i,j}(\Names_{r}
) = \Names_j$. We can then define
\[ P ; Q \eqdef ((P[\phi_{1,2}] \pcomp Q[\phi_{2,3}])\restrict \Names_2 )[
\phi_{1,3}^{-1}] . \]
The picture is
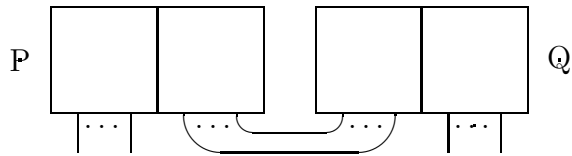
\begin{center}
\begin{picture}(100,60)(10,-15)
\put(-44,20){\framebox(0,0){P}}
\put(-32,0){\framebox(40,40){}}
\put(8,0){\framebox(40,40){}}
\put(68,0){\framebox(40,40){}}
\put(108,0){\framebox(40,40){}}
\put(160,20){\framebox(0,0){Q}}
\put(-22,0){\line(0,-1){15}}
\put(-12,-5){\makebox(0,0){\ldots}}
\put(-2,0){\line(0,-1){15}}
\put(58,0){\oval(80,30)[b]}
\put(58,0){\oval(40,15)[b]}
\put(30,-5){\makebox(0,0){\ldots}}
\put(88,-5){\makebox(0,0){\ldots}}
\put(118,0){\line(0,-1){15}}
\put(128,-5){\framebox(0,0){\ldots}}
\put(138,0){\line(0,-1){15}}
\end{picture}
\end{center}
The key process-algebraic facts are
\[ \begin{array}{lcl}
(P ; Q) ; R & \faileq & P ; (Q ; R) \\
P ; I & \faileq & P \\
I ; P & \faileq & P \\
\lapp{P}{(Q;R)} & \faileq & \lapp{\lapp{P}{Q}}{R} \\
\rapp{(P;Q)}{R} & \faileq & \rapp{P}{\rapp{Q}{R}}
\end{array} \]
\paragraph{Exercise} Verify!

\noindent From these, the fact that $P;Q$ realizes $\Gamma , \Delta$ follows easily.
For example, suppose that $R \negforces \Gamma$. We must show that
$\lapp{R}{P;Q} \posforces \Delta$.
But
\[ P \posforces \Gamma , A \;\; \Longrightarrow \;\; \lapp{R}{P} \posforces A \]
\[ Q \posforces A^{\bot}, \Delta \;\; \Longrightarrow \;\; \lapp{\lapp{R}{P}}{Q}
\posforces \Delta \]
and the result follows since
\[ \lapp{R}{P;Q} \faileq \lapp{\lapp{R}{P}}{Q} . \]
This last step uses the fact that, as already stated, we are taking processes
modulo failures equivalence as our realizers. In fact, all the equivalences
stated so far are valid with respect to much finer notions, in particular
for weak bisimulation. 
\subsubsection{Multiplicatives}
The construction of realizers for the multiplicative rules requires
nothing more than renaming.
\[ \frac{P \posforces \Gamma , A \quad Q \posforces \Delta , B}
{P \otimes Q [\phi ] \posforces \Gamma, \Delta , A \otimes B} 
\quad \quad \quad
\frac{P \posforces \Gamma , A , B}{P[\psi ] \posforces \Gamma , A \Par B} 
\]
for suitable renamings $\phi$, $\psi$.
\paragraph{Exercise} Define $\phi$ and $\psi$, and verify that the required
realizability relations do hold.
\subsubsection{Additives}
\[ \frac{P \posforces \Gamma , A \quad 
Q \posforces \Gamma , B}{\rightbank (\alpha).P
+ \rightbank (\beta ).Q \posforces \Gamma , A \With B} \]
\[ \frac{P \posforces \Gamma , A}{\overline{\rightbank 
(\alpha )}.P \posforces \Gamma ,
A \oplus B} \;\;\;\; \quad 
\frac{Q \posforces \Delta , B}{\overline{\rightbank (\beta )}.Q
\posforces \Delta , A \oplus B}. \]
We can define the process combinators 
\[ \langle P , Q \rangle \eqdef \rightbank (\alpha ) . P + \leftbank (\beta ).Q . \]
\[ \leftbank (P) \eqdef \overline{\leftbank (\alpha )}.P \;\;\;\; 
\rightbank (Q) \eqdef
\overline{\rightbank (\beta )}.Q . \]
The key process algebraic facts we need to show the soundness of the
above rules are
\[ \begin{array}{lcl}
\langle P , Q \rangle ; \leftbank (R) & \faileq & P ; R \\
\langle P , Q \rangle ; \rightbank (S) & \faileq & Q ; S \\
\lapp{R}{\langle P , Q \rangle} & \faileq & \langle \lapp{R}{P} , \lapp{R}{Q} 
\rangle .
\end{array} \]
\paragraph{Exercise}
Verify these equations. Show that the first two hold with respect to weak 
bisimulation, but the third does not. Validating this last equation is
one main reason for working with failures equivalence in this paper.
(Those familiar with the failures-divergences model will note that
this equation only holds in that model on the assumption that $R$
is not the immediately divergent process. It is possible to adapt our
treatment to accomodate divergences. We have not done so here to simplify
the presentation.)
\paragraph{Exercise} Using the above equations, show the soundness
of the realizability assignments for the additive rules.
\paragraph{Exercise}
Work out the realizability assignments for the exponentials $!$ and
$?$ of Linear Logic,
and prove their soundness.
\subsection{Cut-Elimination}
Thus for each sequent proof $\Pi$ in Linear Logic, we can assign a
process term $\lsem \Pi \rsem$. Moreover, we have the following result
expressing the soundness of our assignment with respect to Cut Elimination.
\begin{proposition}
If $ \Pi$ reduces to $\Pi'$ under cut-elimination, then $\lsem \Pi \rsem
\faileq \lsem \Pi' \rsem$.
\end{proposition}
By virtue of these results, we can claim to have modelled Cut-elimination
by process interaction. We can also prove that Cut-elimination
\emph{terminates} in our model.
We outline the argument. For a detailed account in the interaction categories
setting, see \cite{AGN99}.

Firstly, we say that a process $P$ \emph{diverges}, written $P \diverges$,
if there is a sequence $(P_n \mid n \in \Nat )$ with $P = P_0$ and
$ P_n \labarrow{\tau} P_{n+1}$ for all $n \in \Nat$.
We say that $P$ is \emph{convergent}, written $P \converges$,
if it does not diverge.

\noindent Now we define
\[ P \perp Q \iffdef ((P \pcomp Q)\restrict \Names ) \converges . \]
Thus if we ``close the system'', so that $P$ and $Q$ can only interact with
each other, there must be no possible divergences.

\noindent We say that a formula $A$ is \emph{total} if
\[ \forall P \, {\| \!\! -}_{+} A . \; \forall Q \negforces A. \; P \perp Q , \]
and that it is \emph{inhabited} if it has both positive and negative
realizers.
Note that a realizer for a total and inhabited type is convergent.
\begin{proposition}
If $A$ and $B$ are total and inhabited, then  so are
$A^{\bot}$, $A \otimes B$, $A \With B$, $!A$ (and hence also the 
other connectives).
\end{proposition}
This means that if we start from total and inhabited interpretations
of the atomic formulas, then the process we extract from any proof
will be convergent. The key case is Cut, where the totality
condition plays an analogous role to the computability predicate in a
Tait-style proof of strong normalization \cite{Tai,GLT}.
The fact that our process realizers are convergent is
analogous to the fact that a functional realizer extracted by standard 
realizability from a proof say in second-order logic will compute
a total functional. 
\section{The Realizability Category}
We now turn to a more semantic view of process realizability, in the same
general spirit as the by now standard idea of constructing categories
of assemblies or realizability toposes (for which see \textit{e.g.}
\cite{AL,Cro,Lon,AC}).

For each formula $A$, we can define
\[ \begin{array}{lcl}
S_A & = & \{ P \mid P \posforces A \} \\
S_A^{\ast} & = & \{ Q \mid Q \negforces A \}
\end{array} \]
We can then define the ``realizability semantics'' for a formula $A$
as
\[ \lsem A \rsem = (S_A , S_A^{\ast}) . \]
The advantage of this point of view is that we can now abstract to
consider \emph{any} pair $(S, S^{\ast})$ of sets of processes
as a ``type'' of realizers and counter-realizers, and define the action
of the various Linear connectives over these types.
Thus for example
\[ (S, S^{\ast})^{\bot} = (S^{\ast}, S) \]
while $(S, S^{\ast}) \linimpl (T, T^{\ast})$ is the pair
\[ 
(\{ P \mid  \forall Q \in S. \, \lapp{Q}{P} \in T
\conj 
\forall R \in T^{\ast}. \, \rapp{P}{R} \in S^{\ast} \} , 
\{ P \otimes Q \mid  P \in S \conj Q \in T^{\ast} \} )
\]
etc. This idea needs to be refined slightly to yield a satisfactory result. 
In particular, in order to build in a suitable ``modulus of extensionality'',
we shall work with \emph{partial equivalence relations} on processes,
rather than simply sets of processes. Recall that a partial equivalence
relation is a symmetric, transitive relation.

We shall take a type to be a pair $(E, E^{\ast})$ of partial equivalence
relations on processes, where processes are identified up to failures
equivalence. The interpretations of the linear connectives lift to
operations on partial equivalence relations in a straightforward manner.
For example, $E_{A \linimpl  B}$ consists of all pairs $(P, Q)$ such that
\[ 
\forall (R, S) \in E_A . \, 
(\lapp{R}{P}, 
\lapp{S}{Q}) \in E_B  
\conj 
\forall (T, U) \in E_B^{\ast} . \, (\rapp{P}{T},  \rapp{Q}{U}) \in
E_A^{\ast}.
\]
We proceed to define a realizability category $\Cat$. The objects 
\[ A = (E_A, E_{A}^{\ast}) \]
are pairs of partial equivalence relations on processes.
A morphism $f : A \rightarrow B$ is a partial equivalence class $[P]$
of $E_{A \linimpl B}$. This partial equivalence class induces a pair
of maps $(f^{+},f^{-})$, where $f$ maps partial equivalence classes
of $E_A$ to partial equivalence classes of $E_B$, and $f^{-}$ maps
partial equivalence classes of $E_B^{\ast}$ to partial equivalence
classes of $E_A^{\ast}$:
\[ f^{+}([Q]) = [\lapp{Q}{P}], \;\;\;\; f^{-}([R]) = [\rapp{P}{R}]. \]
Conversely, any such pair of maps which is ``tracked'' by a process $P$
in this way determines a unique partial equivalence class of $E_{A \linimpl B}$.

The further structure of the category unfolds as essentially a recapitulation
of our account of realizability for linear proofs. For example, identities and 
composition are realized as for Axiom and Cut. Our constructions for
$\otimes$ and $\linimpl$ give $\Cat$ the structure of a symmetric monoidal
closed category. There is a  duality
\[ \frac{f : A \rightarrow B}{f^{\bot} : B^{\bot} \rightarrow A^{\bot}} \]
where if $f = (f^{+},f^{-})$, $f^{\bot} = (f^{-},f^{+})$. At the process
level, the duality just amounts to interchanging the left-right
partition of the name space
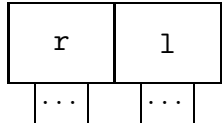
\begin{center}
\begin{picture}(100,50)(0,-15)
\put(8,0){\framebox(40,30){$\rightbank$}}
\put(48,0){\framebox(40,30){$\leftbank$}}
\put(18,0){\line(0,-1){15}}
\put(28,-7.5){\makebox(0,0){\ldots}}
\put(38,0){\line(0,-1){15}}
\put(58,0){\line(0,-1){15}}
\put(68,-7.5){\makebox(0,0){\ldots}}
\put(78,0){\line(0,-1){15}}
\end{picture}
\end{center}
The additive connectives $\With$, $\oplus$ give products and coproducts
in $\Cat$, and in fact $\Cat$ has all (countable) limits and colimits.
$!A$ gives the cofree cocommutative comonoid on $A$.
\begin{proposition} $\Cat$ is a model of Linear Logic.
\end{proposition}

\paragraph{Exercise}
Verify some of this structure. For example, show that $\With$
gives the categorical product in $\Cat$.
\subsection{Quantifiers}
This construction is easily extended to yield a model of second-order
Linear Logic.

A formula $A[X]$ with a second-order propositional variable $X$
can be interpreted as a function $F_A$ on the objects of $\Cat$ in
the obvious fashion. We then define
\[ E_{\forall X. \, A[X]} = \bigcap \{ E_{F_A (B)} \mid B \in \Ob \, \Cat \} \]
\[ E_{\forall X. \, A[X]}^{\ast} = (\bigcup \{ E_{F_A (B)}^{\ast} \mid
B \in \Ob \, \Cat \} )^{+} . \]
The clause for the negative realizers is really for the second-order
existential: 
\[ (\forall X. \, A[X])^{\bot} = \exists X. \, A[X]^{\bot} . \]
The transitive closure is used, since the union of a family of per's need not
be transitive. This ``information loss'' is typical of the behaviour of 
second-order existentials.

First-order quantifiers are handled nicely by value-passing at the
process algebra level. We take the set $\Val$ of values to be the domain
of quantification. The realizability definitions are similar to those for
the additives.
We fix an atomic name $\sigma$.
\[ P \posforces \forall x. \, A \equivdef P \faileq \sigma x.\, Q \conj
\forall v \in \Val . \, Q[v/x] \posforces A[v/x] \]
\[ P \negforces \forall x. \, A \equivdef P \faileq \bar{\sigma}v.R \conj
R \negforces A[v/x] . \]
In terms of operations on partial equivalence relations:
\[ E_{\forall x. \, A} = \{ (\sigma x.P, \sigma x.Q) \mid \forall v \in \Val .
\, (P[v/x], Q[v/x]) \in E_{A[v/x]} \} \]
\[ E_{\forall x. \, A}^{\ast} = \{ (\bar{\sigma}v.P, \bar{\sigma}v.Q) \mid
(P, Q) \in E_{A[v/x]}^{\ast} \} . \]
\subsection{Inductive and Co-inductive types}
Inductive and coinductive types can be canonically interpreted in $\Cat$
as initial $T$-algebras and final $T$-coalgebras for endofuctors
$T : \Cat \rightarrow \Cat$. 
We look at a basic example by way of illustration.
Firstly, we define a unit type $I$ with
\[ E_I = E_I^{\ast} = \{ (0,0) \} . \]
We fix a type $A$, and define a type of $A$-lists by
\[ L = \mu X. \, I \oplus (A \otimes X) . \]
What are the realizers for $L$? The empty list is realized by
$\bar{\alpha}.0$. Given a realizer $P$ for $A$, which may be taken
as realizing the ``value'' $v = [P]$, the unit list $[v]$ is realized
by $\bar{\beta}.(P \otimes \bar{\alpha}.0)$. Inductively, if
an $A$-list $l$ is realized by $Q$, and an $A$-value $v$ is realized by $P$,
then the list $v :: l$ is realized by $\bar{\beta}.(P \otimes Q)$.
Thus we get an inductive definition of the positive realizers for $L$, which
may be compared to that for $?A$. It is worth noting that these lists
are truly \emph{linear}---to ``read'' them (by interacting with the process
realizing the list) is to consume them.

\noindent 
What of the counter-realizers for $L$? These will be the $A$-list consuming 
contexts. Such a context can be defined as follows.
Let $(Q_n \mid n \in \Nat )$ be a family of processes, with
\[ Q_n \negforces \underbrace{A \otimes \cdots \otimes A}_n , \;\; 
(n \in \Nat ) . \]
Define a family $(P_n \mid n \in \Nat )$ by simultaneous recursion:
\[ P_i = \alpha . Q_i + \beta . P_{i+1} \;\; (i \in \Nat ) . \]
Then $P_0$ is a counter-realizer for $L$.
\paragraph{Exercise} Work out the details of this example, to give an
explicit description of the initial $T$-algebra for the endofunctor
\[ T X = I \oplus (A \otimes X) . \]
\paragraph{Exercise} Similarly, analyze the coinductive type of $A$-streams:
\[ S = \nu X. \, I \, \With  \, (A \otimes X). \]
We can use the example of lists to give a useful intuition for the symmetric 
condition on realizers for the linear implication.
Consider a morphism
\[ f : \mathsf{List} (A) \rightarrow \mathsf{List} (B) . \]
This means that we have a realizer $P$ for a pair of maps $(f^{+}, f^{-})$.
In the forwards direction, $P$ induces a function $f^{+}$ mapping
$A$-lists to $B$-lists. In the backwards direction, $P$ induces a
\emph{context-transformer} $f^{-}$ mapping contexts whch consume
$B$-lists to contexts which consume $A$-lists.
E.g. given the definition
\[  \begin{array}{lcl}
\mathtt{f} \; [ \  ] & = & [ \  ] \\
\mathtt{f} \; a :: b :: xs & = & (a+b) :: xs
\end{array} \]
we have the usual function
\[ f^{+}([1, 2]) = [3] \]
and also
\[ f^{-}(\hd ([\cdot ]) = \hd (\hd [\cdot ]) . \]
The context-transformer part of the interpretation of $f$ in our realizability
semantics is \emph{intensional} information about the behaviour of $f$
as an \emph{algorithm}, rather than merely a set-theoretical function.
This opens up the possibility of accurate 
realizability models for non-functional
languages.


\begin{thebibliography}{DeLillo}

\bibitem[Abr94a]{Abr}
S. Abramsky.
Interaction Categories and Communicating Sequential Processes.
In \textsl{A Classical Mind. Essays in Honour of C. A. R. Hoare},
ed. A. W. Roscoe, 1--16. Prentice Hall 1994.

\bibitem[Abr94b]{Abr94}
S. Abramsky.
Proofs as Processes.
\textsl{TCS} vol. 135, 5--9, 1994.

\bibitem[AGN96]{AGN96}
S. Abramsky, S. J. Gay and R. Nagarajan.
Interaction Categories and the Foundations of Typed Concurrent
Programming. In M. Broy, ed. \textsl{Deductive Program Design:
Proceedings of the 1994 Marktoberdorf Summer School}, 35--113.
Springer, 1996.

\bibitem[AGN99]{AGN99}
S. Abramsky, S. J. Gay and R. Nagarajan.
A specification structure for deadlock-freedom of synchronous processes.
\textsl{TCS}, vol. 222, 1--53, 1999.

\bibitem[AC98]{AC}
R. Amadio and P.-L. Curien.
\textsl{Domains and Lambda-Calculi}.
Cambridge University Press, 1998.

\bibitem[AL91]{AL}
A. Asperti and G. Longo.
\textsl{Categories, Types and Structures}.
MIT Press, 1991.

\bibitem[BW99]{BW}
M. Barr and C. Wells.
\textsl{Category Theory for Computing Science. Third Edition}.
Les Publications CRM, Montreal, 1999.

\bibitem[BR83]{BR83}
S. Brookes and W. Rounds.
Behavioural Equivalences induced by Programming Logics.
In \textsl{Proceedings of ICALP `83}, Springer Lecture Notes
in Computer Science vol. 154, 97--108, 1983.

\bibitem[BHR84]{BHR}
S. Brookes, C. A. R. Hoare and A. W. Roscoe.
A Theory of Communicating Sequential Processes.
\textsl{JACM} vol. 31, no. 7, 560--599, 1984.

\bibitem[BR84]{BR84}
S. Brookes and A. W. Roscoe.
An Improved Failures Model for Communicating Processes.
In Springer Lecture Notes in Computer Science vol. 197, 281--305, 1984.

\bibitem[Cro93]{Cro}
R. L. Crole.
\textsl{Categories for Types}.
Cambridge University Press, 1993.

\bibitem[DNH83]{DNH}
R. DeNicola and M. Hennessy.
Testing Equivalence for Processes.
\textsl{TCS} vol. 34, 83--133, 1983.

\bibitem[Dir67]{Dir}
P. A. M. Dirac.
\textsl{Principles of Quantum Mechanics}.
Oxford University Press, 1967.

\bibitem[GLT89]{GLT}
J.-Y. Girard, Y. Lafont and P. Taylor.
\textsl{Proofs and Types}.
Cambridge University Press, 1989.

\bibitem[Hen88]{Hen}
M. Hennessy.
\textsl{Algebraic Theory of Processes}.
MIT Press, 1988.

\bibitem[Hoa85]{Hoa85}
C. A. R. Hoare.
\textsl{Communicating Sequential Processes}.
Prentice Hall, 1985.

\bibitem[Kle45]{Kle}
S. C. Kleene.
On the interpretation of intuitionistic number theory.
\textsl{JSL} vol. 10, 1945.

\bibitem[Lon95]{Lon}
J. R. Longley.
\textsl{Realizability Toposes and Language Semantics}.
Ph.D. thesis, Department of Computer Science,
University of Edinburgh, 1995.
\bibitem[Mil89]{Mil89}
R. Milner.
\textsl{Communication and Concurrency}.
Prentice Hall, 1989.

\bibitem[Ros97]{Ros}
A. W. Roscoe.
\textsl{The Theory and Practice of Concurrency}.
Prentice Hall, 1997.

\bibitem[BS94]{Sch}
U. Berger and H. Schwichtenberg.
Program extraction from classical proofs.
In D. Leivant, ed. \textsl{Logic and Computational Complexity},
Springer Lecture Notes in Computer Science vol. 960, 77--97, 1995.

\bibitem[Sco76]{Sco76}
D. S. Scott.
Data Types as Lattices.
\textsl{SIAM J. on Computing}, vol. 5, 522--587, 1976.

\bibitem[Tai67]{Tai}
W. W. Tait.
Intensional interpretation of functionals of finite type I.
\textsl{JSL} vol. 32, 198--212, 1967.
\end{thebibliography}
\end{document}